\documentclass{emulateapj}
\shorttitle{Stripping in Groups} \shortauthors{Hester}

\begin{document}
\title{Ram Pressure Stripping in Groups: Comparing Theory and Observations}

\author{J. A. Hester\altaffilmark{1}}
\email{jhester@princeton.edu}

\altaffiltext{1}{Jadwin Hall, Princeton University, Princeton, NJ,
08544}

\begin{abstract}
Ram pressure stripping may be the dominant mechanisms driving the
evolution of galaxy colors in groups and clusters.  In this paper,
an analytic model of ram pressure stripping is confronted with
observations of galaxy colors and star formation rates in groups
using a group catalog drawn from the Sloan Digital Sky Survey. An
observed increase in the fraction of galaxies residing on the red
sequence, the red fraction, with both increasing group mass,
$M_{gr}$, and decreasing satellite luminosity, $L_{sat}$, is
predicted by the model.  The size of the differences in the red
fraction can be understood in terms of the effect of the scatter
in satellite and cluster morphologies and satellite orbits on the
relationship between $M_{gr}$ and $L_{sat}$ and the stripped gas
fraction. Observations of the group galaxies' H$\delta$ and
4000\AA~break spectral measures and a comparison of the
distribution of $SFR/M_{\ast}$ for star forming galaxies in the
groups and in isolation both indicate that the color differences
observed in the groups are the result of slowly declining SFRs, as
expected if the color change is driven by stripping of the outer
H~\textsc{i} disk.
\end{abstract}

\keywords{galaxies: evolution --- galaxies:clusters:general}

\section{INTRODUCTION}\label{intro}
Galaxies in clusters have earlier type morphologies, redder
colors, and lower star formation rates (SFRs) than galaxies in the
field of similar luminosity~\citep{Gomez et al, Goto et al,
Dressler}. Galaxies in groups and clusters formed both earlier and
in denser environments than their field counterparts.  Cluster
galaxies may therefore naturally appear older than field galaxies.
In addition, group and cluster galaxies are more likely to have
undergone major mergers~\citep{GKK} and are susceptible to
harassment~\citep{MLK} and interactions with the ICM~\citep{Gunn
Gott}. These interactions with the environment work to transform
blue, late-type galaxies into red, early-type galaxies.

Several recent observational studies have explored the
relationship between galaxy properties and environment.
\citet{Blanton et al. 05b} and~\citet{Hogg et al. 03} study the
dependence of environment on galaxy properties and find that
bright, red, and concentrated galaxies prefer to reside in
environments with higher densities. Large galaxy surveys have
revealed strong bimodalities in both stellar properties; color,
SFR, 4000\AA~break; and structural properties; concentration,
central surface brightness, $\mu_0$,~\citep{Baldry et al. 04,
Blanton et al. 03a, Kauffmann et al. 03a, Li et al. 06, Strateva
et al}.   As the local density increases the fraction of galaxies
found in the red and high concentration modes
increases~\citep{Kauffmann et al. 04}, and galaxies in the red or
high concentration modes are more clustered than galaxies in the
blue or low concentration modes~\citep{Li et al. 06}.

Galaxy properties are strongly related to each other as well as to
the local environment.  Bivariate distributions of $M_i$, $g-r$
color, central i-band surface brightness, $\mu_i$, and the i-band
Sersic index, $n_i$, are shown in~\citet{Blanton et al. 05b}.
Galaxies tend to be either blue with low concentrations or red
with high concentrations.  In addition, the average 4000\AA~break,
g-r color, concentration, and $\mu_0$ all monotonically increase
with stellar mass~\citep{Kauffmann et al. 03b}.  Understanding the
effect of environment on galaxy evolution requires studying the
effect of environment on both the distribution of galaxy
properties and on the relationships between galaxy properties.

Three separate studies show that the relationship between color
and environment is stronger than the relationship between
concentration and environment. \citet{Blanton et al. 05b}
demonstrate that the observed relationship between $n_i$ and
environment is reproduced when galaxies are assigned new local
densities based on their $g-r$ colors and i-band magnitudes,
$M_i$. However, the observed relationship between color and
environment cannot be reproduced by assigning local densities
based on $n_i$ and $M_i$. \citet{Kauffmann et al. 03b} show that
the relationship between concentration and stellar mass depends on
the local environment only weakly and only at low stellar masses,
$M_{\ast}<3\times10^{10}M_{\odot}$. In contrast the relationship
between the 4000\AA~break or the SFR and the stellar mass shows a
strong dependence on the local environment. \citet{Li et al. 06}
study the correlation function, in bins of luminosity, for
galaxies belonging to the different galaxy modes. Galaxies that
are red, have large 4000\AA~breaks, high concentrations, or bright
$\mu$ all have enhanced correlation functions on scales less than
5Mpc. However, for galaxies that are red and have large breaks,
the enhancement is larger and extends to larger scales.

If environment causes galaxy colors to evolve, then the timescale
on which evolution occurs is a clue to the processes responsible.
\citet{Kauffmann et al. 04} and~\cite{Balogh et al. 04} both study
the timescale over which galaxies' colors evolve and come to
different conclusions.  \citet{Kauffmann et al. 04} compare Sloan
Digital Sky Survey (SDSS) spectra to stellar evolution models and
observe that the absence of galaxies with strong H$\delta$ for
their 4000\AA~break strengths argues against the predominance of
any process that truncates star formation on timescales less than
a Gyr. In contrast,~\citet{Balogh et al. 04} use observations of
$H\alpha$ equivalent widths to argue against slowly evolving SFRs.
While they observe an increase in the fraction of non-star forming
galaxies in high density environments, they do not see a downward
shift in the $W(H\alpha)$ distributions for either actively star
forming galaxies or galaxies with blue colors. These observations
indicate that galaxies pass quickly from the star forming to the
non-star forming populations.

Ram pressure stripping of galaxies in groups and clusters is one
process that can truncate star formation and redden galaxies.
\citet{P1} (Hereafter, Paper 1) presents a model of ram pressure
stripping that predicts the fraction of the gas that is stripped
from the outer H~\textsc{i} disk and the hot galactic halo. This
fraction depends primarily on the ratio of the galaxy's mass to
the mass of the group or cluster in which it orbits and secondly
on several descriptive model parameters.

In this paper, the importance of ram pressure stripping is
examined by combining the predictions of the model presented in
Paper 1 with observations of galaxy groups in the SDSS. The group
catalog used here was assembled by A. Berlind and is presented
in~\citet{Berlind et al.}.  It is reviewed briefly in
\S~\ref{SDSS}. The predictions of the model for the galaxy and
group masses of interest are presented in \S~\ref{Model}. In
\S~\ref{stripping and sfr}, the timescale on which the SFR
declines after a galaxy is stripped of the gas outside its inner
star forming disk is discussed. In \S~\ref{r and d}, observations
of the groups are compared to past observations and to the model
of ram pressure stripping.  The timescale of the decline in the
SFR in the groups and the role ram pressure stripping may play in
determining general trends in galaxy properties are also examined.
\S~\ref{conclusion} concludes.

\section{THE GROUP CATALOG}
\label{SDSS} The Sloan Digital Sky Survey (SDSS)~\citep{York} is
conducting an imaging and photometric survey of $\Pi$ $sr$ in the
northern hemisphere as well as three thin slices in the southern
hemisphere.  Observing is done using a dedicated 2.5m telescope in
Apache Point, NM.  The telescope operates in drift scan mode and
observes in five bandpasses~\citep{Fukigita et al.}.  Magnitude
calibration is carried out using a network of standard
stars~\citep{Smith}. Three sets of spectroscopic targets are
selected automatically, the main galaxy sample, the luminous red
galaxy sample, and the quasar sample~\citep{Straus et al.,
Eisenstein et al., Richards et al.}. Objects in the main galaxy
sample have Petrosian magnitudes $r'<17.7$ and are classified as
extended.  Magnitudes are corrected for galactic extinction using
the reddening maps of ~\citet{Schlegel Finkbeiner Davis} prior to
selection. Spectroscopy is taken using a pair of fiber-red
spectrographs, and targets are assigned to fibers using an
adaptive tiling algorithm~\citep{Blanton et al. 03d}. Data
reduction for the SDSS is done using a series of automated
pipelines~\citep{Hogg et al. 01, Ivezic et al., Lupton et al.,
Pier et al., Smith}.

The catalog used here is a volume limited sample drawn from the
NYU Value Added Galaxy Catalog (NYU-VAGC)~\citep{Blanton et al.
05a}. The sample goes down to $M_{r}<-19.0$ and has a redshift
range of 0.015-0.068. The group finding algorithm is described in
detail in~\citet{Berlind et al.}. The group finder is a friends of
friends algorithm with two linking lengths, one for projected
distances and one in red shift space. Linking lengths are chosen
such that the multiplicity function, richness, and projected size
of recovered groups from simulations projected into redshift space
are unbiased measures. The linking lengths are also chosen to
maximize the number of groups recovered and minimize the number of
spurious groups. The velocity dispersions of the recovered groups
are systematically low because the group finder does not link the
fastest moving group members to the group. However, as the fastest
group members are also those most likely to be stripped, not
including these galaxies biases against observing the signature of
ram pressure stripping.

Virial masses and radii for the groups are determined by assuming
a monotonic relationship between group mass and richness and
matching a $\Lambda$CDM mass function to the measured group
luminosity function~\citep{Berlind et al.}. This assumes no
scatter in the relationship between mass and richness. In this
paper, galaxies are grouped by virial mass, and the assumption
that there is no scatter should not be important. The absolute
magnitudes used here are k-corrected and corrected for passive
evolution to $z=0.1$~\citep{Blanton et al. 03b, Blanton et al.
03c}. Membership in the red sequence is defined using the $M_r$
dependent color cut presented in~\citet{Li et al. 06};
$g-r>-0.788-0.078M_r$. The i-band Sersic index, $n_i$, is defined
as $I_i(r)=A\exp\left[-(r/r_0)^{(1/n_i)}\right]$ and its
measurement is discussed in~\cite{Blanton et al. 03c}. The
$H\delta$ and the 4000\AA~break are those measured
by~\citet{Kauffmann et al. 03a}. The measurements of the SFR are
from~\citet{Brinchmann et al} and are normalized to the galaxies'
stellar masses. The affect of the changing physical size of the
spectroscopic fiber's 3 arcsec diameter is corrected for in
$SFR/M_{\ast}$ by assuming that the relationship between color and
$SFR/M_{\ast}$ is constant throughout the disk. This correction
may bias against observing galaxies with blue colors and low
$SFR/M_{\ast}$ and is important for galaxies at the low redshifts
of this sample.\footnote{H$\delta$, 4000\AA~break, $M_{\ast}$, and
$SFR/M_{\ast}$ are available from
http://www.mpa-garching.mpg.de/SDSS/}

\section{RAM PRESSURE STRIPPING}
\citet{Gunn Gott} proposed ram pressure striping to explain the
observed absence of gas rich galaxies in clusters. Galaxies in
clusters feel an intracluster medium (ICM) wind that can overcome
the gravitational attraction between the stellar and gas disks and
strip the gas disk. They introduced the following condition to
estimate when this occurs;
\begin{equation}\label{eqn rps}
\rho_{ICM}v_{sat}^{2} < 2 \pi G\sigma_{\ast}\sigma_{gas}.
\end{equation}
The left-hand side is a ram pressure, where $\rho_{ICM}$ is the
density of the ICM and $v_{sat}$ is the orbital speed of the
satellite. The right-hand side is a gravitational restoring
pressure where $\sigma_{\ast}$ and $\sigma_{gas}$ are the surface
densities of the stellar and gas disks respectively. Using this
condition they concluded that spirals should lose their gas disks
when they pass through the centers of clusters.

Galaxies in nearby clusters are observed to be deficient in
H~\textsc{i} and to have truncated gas disks when compared to
field galaxies of similar morphology and optical
size~\citep{Bravo-Alfaro et al 2000, Cayatte et al., Giovanelli
and Haynes 1983, Solanes et al 2001}. In addition, asymmetric
extra-planar gas that appears to have been pushed out of the disk
is observed in several Virgo spirals~\citep{Kenney et al 2004,
Kenney and Koopmann, Kenney van Gorkom Vollmer 2004}.  These
observations can be explained by ram pressure stripping, and
observing an undisturbed stellar disks accompanied by extra-planar
gas is a strong indication that the gas disk is interacting with
the ICM. Ram pressure stripping has also been repeatedly observed
in simulations of disk galaxies in an ICM wind~\citep{Abadi Moore
Bower, Marcolini Brighenti D'Ercole, Quilis Moore Bower, Roediger
and Hensler, Schulz and Struck 2001}.

Paper 1 uses an analytical model of ram pressure stripping to
explore the range of environments in which stripping can occur and
the galaxy masses that are susceptible to stripping in each
environment.  It focuses on the H~\textsc{i} disk beyond the
stellar disk, and stripping of gas from within the stellar disk is
not modeled.  The gravitational restoring pressure is found by
placing a flat H~\textsc{i} disk in a gravitational potential
consisting of a dark matter halo, a stellar disk, and a stellar
bulge.  The ram pressure is determined by letting the satellite
orbit in an NFW potential through a $\beta$ profile ICM. The gas
fraction that a galaxy is striped of is found to depend on the
ratio of the satellite mass to the group mass, $M_{sat}/M_{gr}$,
and the values of several descriptive parameters. Observations of
stripped spirals in clusters compare well with the model's
predictions for large $M_{gr}$ and $M_{sat}$. Paper 1 concludes
that many galaxies, particularly low-mass galaxies, can be
stripped of a substantial fraction of their outer H~\textsc{i}
disks in a wide range of environments. The specific model
predictions for the SDSS group catalog are given below and the
possible effects of stripping the outer gas are discussed.

\subsection{Model Predictions}\label{Model}
In this section, the analytic model developed in Paper 1 is used
to predict trends for galaxy colors in the SDSS groups.

The model predicts that the extent of stripping depends on the
ratio $M_{sat}/M_{gr}$.  However, two separate trends should in
fact be observed, one with $M_{gr}$ at fixed $M_r$ and one with
$M_r$ at fixed $M_{gr}$. This is mainly because the average ICM
density of groups decreases as $M_{gr}$ decreases, and the average
gas fraction in the disk increases as $M_{sat}$ decreases, which
shifts the values of $M_{sat}/M_{gr}$ for which stripping
occurs~\citep{Sanderson Ponman, Swaters et. al.}. In addition
there is a strong color magnitude relation seen in both groups and
the field.  Therefore, it is necessary to in fact observe the
change in the difference between the red fraction in the groups
and in isolation with $M_r$ rather than the red fraction itself.

Observing a color or SFR trend due to ram pressure stripping
requires a group catalog with both an appropriate range of group
and galaxy masses and a sufficient number of galaxies and groups.
A useful group catalog is one in which the effectiveness of
stripping varies so that trends in $f_r$ with $M_{gr}$ and $M_r$
are expected. The groups in the catalog have masses between
$10^{12}$ and $10^{15}M_{\odot}$ and the galaxies have
$-19>M_r>-22$.  In Tables~\ref{tbl50} -~\ref{tbl90} the model's
predictions for the satellite mass, $M_{sat}$, and absolute
magnitude, $M_r$, at which a galaxy is stripped of $\approx
50,~80,~90\%$ of its H~\textsc{i} disk are given for groups of
masses $M_{gr} = 10^{13, 13.5, 14,14.5} M_{\odot}$.  The model
galaxy parameters are those from Paper 1 for a large spiral. The
ICM parameters for $M_{gr}=10^{13}M_{\odot}$ are from the low-mass
group model. The ICM parameters for $M_{gr}=10^{13.5, 14,14.5}
M_{\odot}$ are from the middle-mass cluster model. The values in
Tables 1-3 assume a galaxy orbiting inclined to the ICM wind that
is stripped of an intermediate gas fraction between those expected
for a galaxy traveling face-on or edge-on to the wind.  Total mass
to light ratios in the r-band of 40, 65, and 90 are used to
convert $M_{sat}$ to $M_r$. These are a factor 10-15 higher than
the stellar mass to light ratios of 4-6 found for the R-band
by~\citet{Maraston}. In Tables 1-3 galaxies with $M_r<-19$ are
highlighted.  The predictions of the model are not exact, but they
do demonstrate that the range of $M_{gr}$ and $M_r$ in the group
catalog is appropriate for this project. The high-mass groups
should contain many stripped galaxies, but the low-mass groups are
capable of stripping few of the galaxies in the sample. Though the
range in $M_{sat}$ is not as great as the range in $M_{gr}$, in
higher mass groups, galaxies with $M_r=-21$ and $M_r=-19$ are
stripped of different gas fractions.

\begin{deluxetable}{ccccc}
\tabletypesize{\scriptsize} \tablecaption{Gas fraction = 0.5}
\tablewidth{0pt} \tablehead{ \colhead{$M_{v,gr}$} &
\colhead{$M_{v,sat}$} & \colhead{$MtoL = 65$} & \colhead{$MtoL =
40$} & \colhead{$M/L = 90$}} \startdata

13.0  &    11.2   &  \textbf{-19.8}   &  \textbf{-20.3}   &  \textbf{-19.4}\\
13.5  &    11.7   &  \textbf{-21.1}   &  \textbf{-21.6}   &  \textbf{-20.7}\\
14.0  &    12.2   &  \textbf{-22.3}   &  \textbf{-22.8}   &  \textbf{-22.0}\\
14.5  &    13.0   &  \textbf{-23.3}   &  \textbf{-23.8}   &  \textbf{-23.0}\\
\enddata
\label{tbl50}
\end{deluxetable}

\begin{deluxetable}{ccccc}
\tabletypesize{\scriptsize} \tablecaption{Gas fraction = 0.8}
\tablewidth{0pt} \tablehead{ \colhead{$M_{v,gr}$} &
\colhead{$M_{v,sat}$} & \colhead{$MtoL = 65$} & \colhead{$MtoL =
40$} & \colhead{$M/L = 90$}} \startdata

13.0  & 10.9  & -18.2     & -18.7   &  -17.8 \\
13.5  & 11.4  &   \textbf{-19.5}   &  \textbf{-20.0}  &   \textbf{-19.1}\\
14.0  & 11.9  &   \textbf{-20.7}   &  \textbf{-21.2}  &   \textbf{-20.4}\\
14.5  & 12.6  &   \textbf{-21.7}   &  \textbf{-22.2}  &   \textbf{-21.3}\\
\enddata
\label{tbl80}
\end{deluxetable}

\begin{deluxetable}{ccccc}
\tabletypesize{\scriptsize} \tablecaption{Gas fraction = 0.9}
\tablewidth{0pt} \tablehead{ \colhead{$M_{v,gr}$} &
\colhead{$M_{v,sat}$} & \colhead{$MtoL = 65$} & \colhead{$MtoL =
40$} & \colhead{$M/L = 90$}} \startdata

13.0  &    10.2   &  -17.1  &  -17.7   &  -16.8 \\
13.5  &    10.7   &  -18.4  &  \textbf{-18.9}   &  -18.1 \\
14.0  &    11.2   &  \textbf{-19.6}  &   \textbf{-20.2}  & \textbf{-19.3}  \\
14.5  &    12.0   &  \textbf{-20.6}  &   \textbf{-21.1}  &   \textbf{-20.3}\\

\enddata
\label{tbl90}
\end{deluxetable}

The model also predicts the extent of stripping of the hot
galactic halo. The galactic halo is modeled by placing gas at the
virial temperature of the galaxy's dark matter halo in hydrostatic
equilibrium with an NFW potential.  It is assumed that the
galactic halo is stripped down to the radius at which the ram
pressure equals the thermal pressure. The galaxies in this sample
can only maintain $\le0.5$\% of their mass in the galactic halo
without the gas cooling rapidly. Using this mass fraction, the
model predicts that the galaxies in the group sample have more
than 70\% of this gas stripped in a $10^{13.5}M_{\odot}$ group and
the entire halo is stripped in any group with a mass above
$10^{14}M_{\odot}$. This is an upper limit on the hot halo gas
these galaxies can retain in the absence of fresh in-falling gas.
Therefore, it is assumed for the rest of the paper that the
galactic halo is stripped for all galaxies in the sample, and that
any observed trends in $f_r$ are due to ram pressure stripping of
the H~\textsc{i} disk.

The size of the galaxy catalog needed to detect stripping is
determined by the scatter in the relationship between the fraction
of the H~\textsc{i} disk mass that is stripped, $M_{str}$, and the
satellite galaxy mass, $M_{sat}$.  The scatter can be thought of
in terms of a distribution of effective masses, $M_{eff}$, at each
physical satellite mass $M_{sat}$. The effective mass is defined
such that all satellites with the same effective mass are stripped
of the same fraction of their H~\textsc{i} disk. This scatter is
substantial. As shown in Paper 1, a $10^{11}M_{\odot}$ galaxy can
be stripped as much as a $10^{10}M_{\odot}$ galaxy or as little as
a $10^{12}M_{\odot}$ galaxy.  The scatter is mainly due to
differences in the galaxies' orbits, stellar and H~\textsc{i} disk
scale lengths, and in the density and extent of the groups' ICM.
Paper 1 discusses the number of groups and galaxies a sample must
contain in order to observe systematic changes in $M_{str}$ with
$M_{sat}$ and $M_{gr}$. However, the relationship between
$M_{str}$ and color is not simple. Therefore, because the
connection between $M_{str}$ and residing on the red sequence is
probably more direct, the red fraction will be studied rather than
average galaxy colors. Galaxies that have been stripped of more
than a critical gas fraction, $M_c$, should eventually join the
red sequence. In this picture, the fraction of blue disk galaxies
entering groups that join the red sequence is the fraction with
$M_{str}>M_{c}$, and the fraction of galaxies with a given
$M_{sat}$ that join the red sequence is intimately related to the
distribution of $M_{eff}$ at $M_{sat}$.  If the scatter in
$M_{eff}$ is large compared to the difference in $M_{sat}$ or
$M_{gr}$ between two sub-samples, the difference in the red
fraction between the sub-samples should be small. For this sample,
the scatter in $M_{eff}$ is similar range in $M_{sat}$, but
smaller than the range in $M_{gr}$.

To get a rough estimate of the necessary catalog size, the number
of galaxies needed to observe a difference in the red fraction of
0.1 is determined here.  As errors in the colors are small in
comparison, it is assumed that the scatter in the observed red
fraction is given by $\sigma_{fr}^2 = f_{rt}(1-f_{rt})/N$, where
$f_{rt}$ is the true red fraction.  In the case that two $f_{rt}$
are close to 0.5, approximately 400 galaxies are needed in each
sub-sample to observe a difference in the observed red fractions
between them of 0.1 with 3$\sigma$ confidence. The scatter in
$M_{eff}$ is due to both variations in orbital and satellite
parameter values, which vary within a single group, and to
variations in the ICM parameter values, which vary across groups.
Therefore, this estimate is only valid for a sample that includes
a large number of groups. The SDSS group catalog contains 2700
groups, 15400 group galaxies, and 22500 isolated galaxies. It is
large enough to place 400 galaxies into reasonable bins in $M_r$
and $M_{gr}$ and has $\langle N_{gal}/N_{gr}\rangle\approx 5.5$.
In other words, the catalog is large enough to detect differences
in the red fraction as small as 0.1 with reasonable significance.
This demonstrates why observations of colors and SFRs in the SDSS
groups are being used to search for the signal of ram pressure
stripping rather than the more obvious approach of using
H~\textsc{i} observations. Potentially observing a trend due to
ram pressure stripping requires observing thousands of galaxies,
preferably across a wide range of satellite and group masses, with
uniform determinations of galaxy mass or luminosity and group
mass.

The catalog of SDSS groups used for this project covers a range of
$M_{gr}$ and $M_r$ in which stripping should occur and across
which the degree of stripping should vary. The fraction of
galaxies that belong to the red sequence will be focused on rather
than average galaxy colors.  This is mainly because residence in
the red sequence and ram pressure stripping should be related in a
more straight forward manner.  It also simplifies comparisons
between this project and others that use the galaxy color
bimodality.  The group catalog should be large enough to observe
the signal of ram pressure stripping. The range in $M_{gr}$ is
larger than that in $M_{sat}$.  Therefore, the change in $f_r$
across the $M_{gr}$ range should be larger.

\subsection{Gas Loss and the SFR}\label{stripping and sfr}
The model of ram pressure stripping presented in Paper 1 addresses
the stripping of the outer H~\textsc{i} disk. Galaxies can be
stripped of their inner gas disk.  However, this probably only
happens to the smallest galaxies or in the highest density
environments.  As is seen in Paper 1 and reviewed above, ram
pressure stripping of the outer gas disk occurs for a wide range
of satellite masses and environments.  In this paper, the colors
of galaxies, rather than H~\textsc{i} observations, are used to
confront the predictions of the disk stripping model. It is
therefore assumed that stripping of the outer H~\textsc{i} disk
affects galaxies' future SFRs.  This section discusses the
likelihood that the outer gas disk fuels future star formation and
the time scale on which the SFR should decline if this gas is
removed.

While disk galaxies are currently forming stars from the gas in
their inner disks, they cannot continue to form stars at their
current rate for longer than a few Gyr unless this gas is
replenished.  Outside of groups star formation can be sustained by
in-fall of new gas into and the continual cooling of the hot
galactic halo.  However, in almost any group environment, most
galaxies are stripped of their hot galactic halo gas and
experience no new in-fall. If the loss of this gas were
responsible for the majority of the reddening seen in groups, then
all but the brightest galaxies would be uniformly reddened in any
group with an ICM.  This is seen for semi-analytic models of
galaxy colors which include galactic halo stripping and ignore the
extended H~\textsc{i} disk~\citep{Weinmann06}.  In this scenario
the SFR declines slowly as the galaxy consumes its inner gas disk.

\begin{figure*}[t]\center
\epsscale{0.95}\plotone{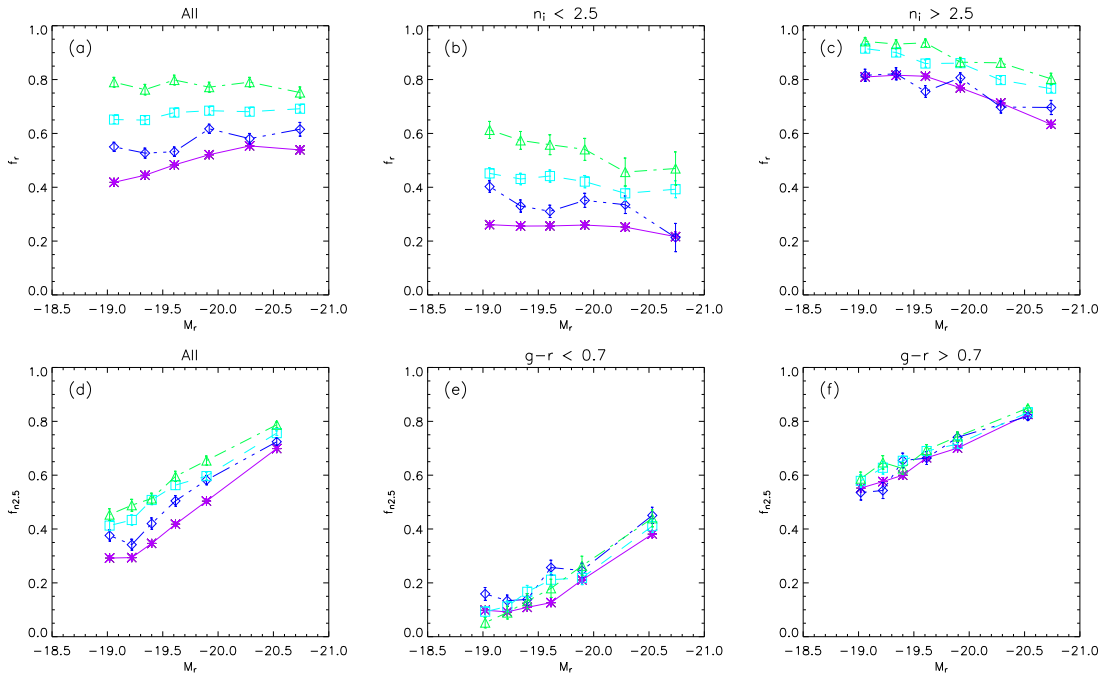} \caption{Fraction of
galaxies that belong to the red sequence, $f_r$, and the fraction
with $n_i>2.5$, $f_{n2.5}$, vs $M_r$ for different group masses,
$M_{gr}$. \textit{Stars/solid}: Isolated galaxies, that is
galaxies not in a group. \textit{Diamonds/dash-double-dot}:
Galaxies in groups with $log(M_{gr}(M_{\odot}))<13$.
\textit{Squares/dash}: $13<log(M_{gr}(M_{\odot}))<14$.
\textit{Triangles/dash-dot}: $13.5<log(M_{gr}(M_{\odot}))<14.5$.
(a): $f_r$ for all galaxies. (b)((c)): $f_r$ for galaxies with
$n_i<2.5$ ($n_i>2.5$). (d): $f_{n2.5}$ for all galaxies. (e)((f)):
$f_{n2.5}$ for galaxies with $g-r<0.7$ ($g-r>0.7$). Note that
galaxies in groups have both a higher $f_r$ and a higher
$f_{n2.5}$ than isolated galaxies, in agreement with previous
studies of the relationship between color and $n_i$ and
environment.  In addition, $f_r$ increases with $M_{gr}$ for all
$n_i$.  This increase is expected if ram pressure stripping
determines the group galaxies SFRs and colors, but not if SFRs in
groups are set by the stripping of the galactic halo
alone.}\label{red fraction}
\end{figure*}

Star formation may also be fed by the inflow of gas from the outer
H~\textsc{i} disk.  Inflow is expected if there is any viscosity
in the disk. In-falling gas should often join the outer
H~\textsc{i} disk, and feeding star formation with a viscous disk
can be used to form exponential stellar profiles~\citep{Bell, Lin
Pringle a, Lin Pringle b}. Observationally,~\citet{Gavazzi} find
that late-type galaxies that are moderately deficient in
H~\textsc{i} as compared to galaxies with the same morphology and
optical size also have low H$\alpha$ equivalent widths for their
morphology. If star formation is fed by gas in the outer gas disk,
then galaxies in groups that retain this gas will be able to
continue forming stars while galaxies that do not will experience
a decline in their SFRs.  In this case galaxies' SFRs and colors
should depend the effectiveness of ram pressure stripping.

When the gas disk is stripped down to the radius at which star
formation is occurring, the time scale over which star formation
declines is determined by the rate at which star formation
consumes the gas disk.  The timescale for the decline in the star
formation rate can be defined as
$t_c\equiv-\Sigma_{SFR}/\dot{\Sigma}_{SFR}$, where $\Sigma_{SFR}$
is the surface density of star formation.  If $\Sigma_{SFR}$ is
related to the gas surface density, $\Sigma_H$, by a local Schmidt
law, $\Sigma_{SFR}\propto\Sigma_{H}^{\alpha}$, then
$t_c\approx\Sigma_{H,0}/(\alpha\Sigma_{SFR,0})$. \citet{Kennicutt}
found that
$\Sigma_{SFR}=2.5\times10^{-4}(\Sigma_H/1M_{\odot}pc^{-2})^{1.4}~M_{\odot}~yr^{-1}~kpc^{-2}$.
For $\Sigma_H$ of order a few $M_{\odot}~pc^{-2}$, this relation
gives $t_c$ of order a Gyr.  If star formation is fed by inflow in
the H~\textsc{i} disk, galaxies in groups that retain a
significant portion of their H~\textsc{i} disk can continue to
form stars.  Galaxies that are stripped of their gas disks will
experience a slow decline in their SFRs with a timescale of order
a Gyr.

\section{RESULTS AND DISCUSSION}\label{r and d}
In this section the properties of the group galaxies are compared
to past observations of the relationship between galaxy properties
and environment and to the model of ram pressure stripping
presented in Paper 1.  The timescale of the star formation decline
occurring in the groups is also examined.  In addition, the role
of ram pressure stripping in determining more general trends in
galaxy properties is discussed.

\subsection{Comparison with Other Observation and with the Model}

\begin{figure*}[t]\center
\epsscale{0.95}\plotone{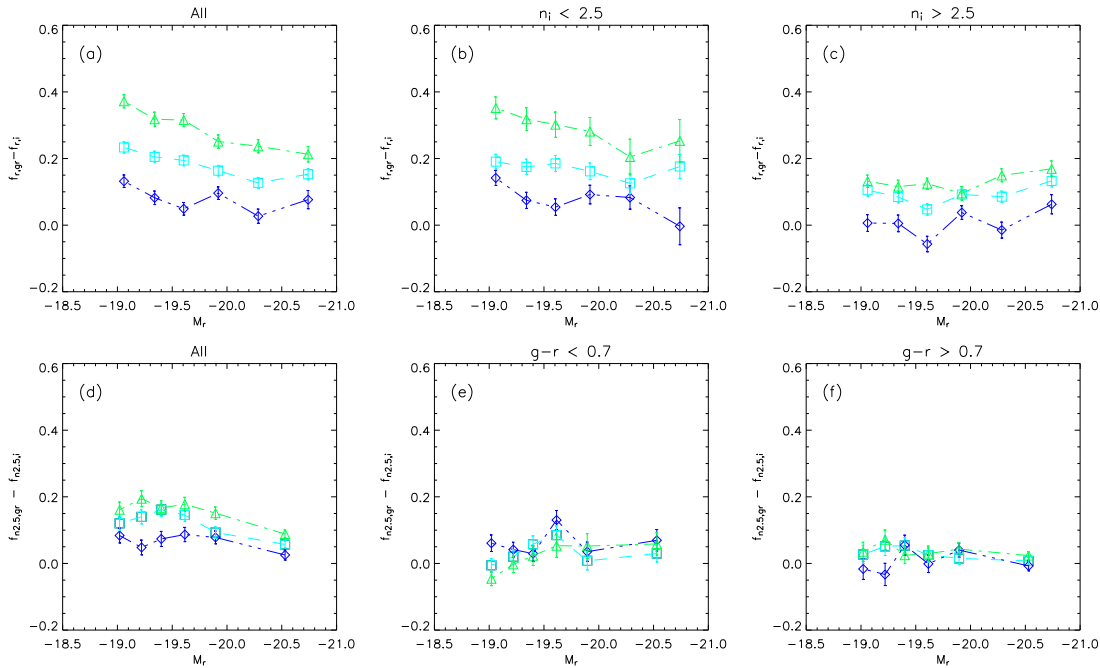} \caption{Difference between
the galaxies in the groups and in isolation in the red fraction,
$f_{r,gr}-f_{r,is}$, and in the fraction of galaxies with
$n_i>2.5$, $f_{n2.5,gr}-f_{n2.5,is}$ for different $M_{gr}$.
Symbols and lines are the same as in Figure~\ref{red fraction}.
(a): $f_{r,gr}-f_{r,is}$ for all galaxies. (b)((c)):
$f_{r,gr}-f_{r,is}$ for galaxies with $n_i<2.5$($n_i>2.5$)., (d):
$f_{n2.5,gr}-f_{n2.5,is}$ for all galaxies. (e)((f)):
$f_{n2.5,gr}-f_{n2.5,is}$ for galaxies with $g-r<0.7$($g-r>0.7$).
Note that an excess of red galaxies is observed for all $n_i$, but
that no excess of galaxies with $n_i>2.5$ is seen in either the
blue or red sub-populations. This asymmetry has been observed
before in studies based on local over-densities and conditional
correlation functions.  Also observe the decrease in the change in
$f_r$ with decreasing $M_r$ (increasing $M_{sat}$) for both all
galaxies and for galaxies with $n_i>2.5$ in the highest $M_{gr}$
bin.  This trend can be understood in terms of a model in which
the SFRs in groups are determined by the effectiveness of ram
pressure stripping.  A possible reason that this trend is not
observed in the $n_i<2.5$ sub-population for the lower two
$M_{gr}$ bins is given in the text.}\label{red fraction
difference}
\end{figure*}

\begin{figure*}[t]\center
\epsscale{0.95}\plotone{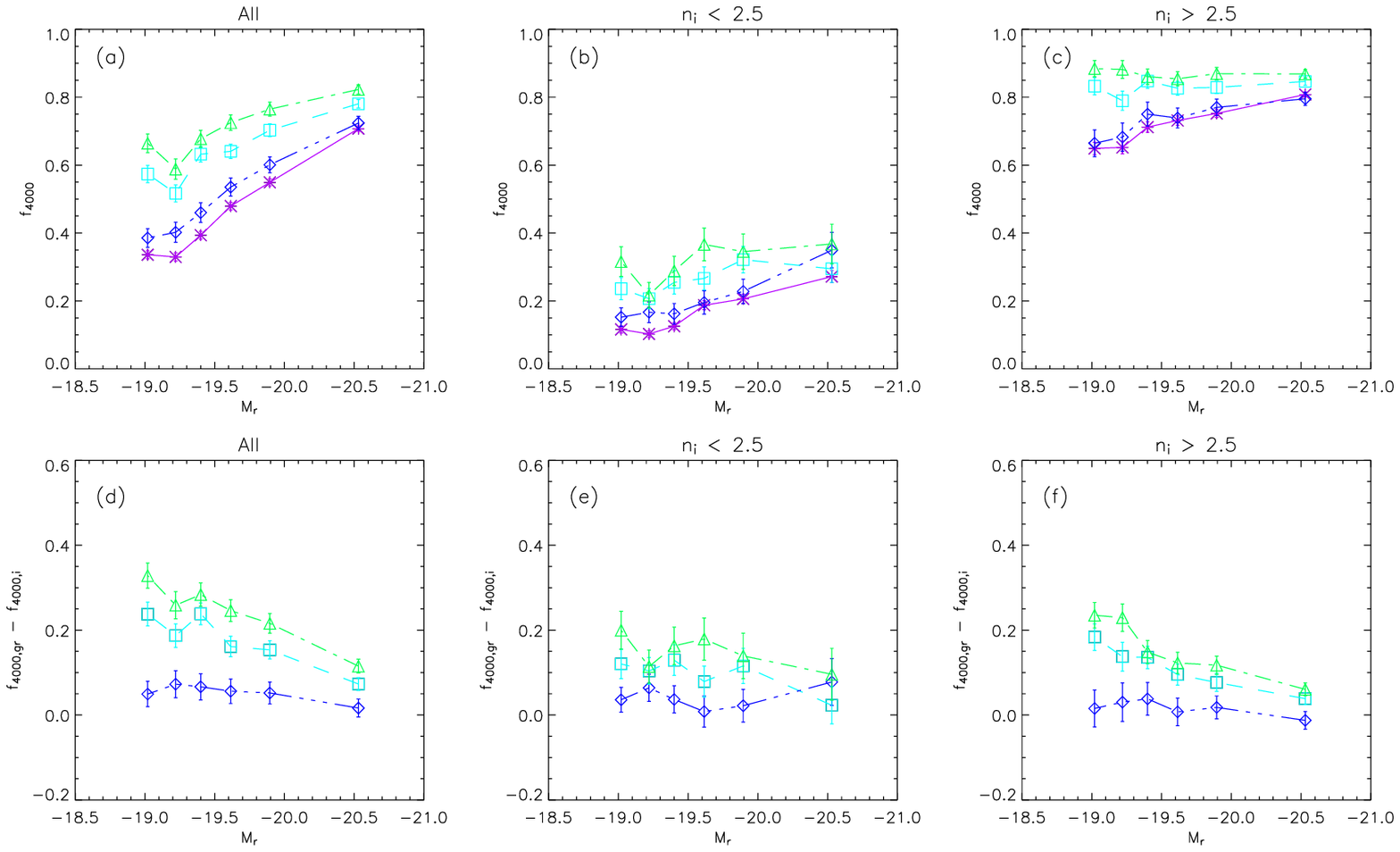} \caption{Fraction of
galaxies with 4000\AA~break measure greater than 1.5, $f_{4000}$,
vs $M_r$ for different $M_{gr}$. Symbols and lines are the same as
in Figure~\ref{red fraction}. (a): All galaxies. (b)((c)):
Galaxies with $n_i<2.5$ ($n_i>2.5$). (d): Difference in $f_{4000}$
between galaxies in the groups and in isolation for all galaxies.
(e)((f)): Difference for galaxies with $n_i<2.5$ ($n_i>2.5$). The
trends in $f_r$ seen in Figures~\ref{red fraction}a-c and~\ref{red
fraction difference}a-c are also seen in the 4000\AA~break. Some
of the differences between color and 4000\AA~break are likely due
to the size of the spectroscopic fiber. As the size of the galaxy
increases, the fraction of the light in the fiber coming from the
central region increases.} \label{break fraction}
\end{figure*}

\begin{figure}[b]
\epsscale{.7} \plotone{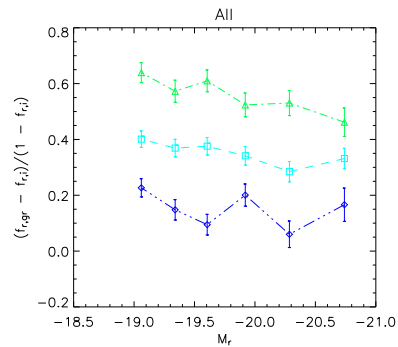}\caption{Difference between
the red fraction in the groups, $f_{r,gr}$ and the red fraction in
isolation, $f_{r,is}$, normalized by the blue fraction in the
field for each bin, $(f_{r,gr}-f_{r,is})/(1-f_{r,is})$. Symbols
and lines are the same as in Figure~\ref{red fraction}. The
normalized difference is an estimate of the percentage of the
in-falling blue galaxies that join the red sequence.  This plot
emphasizes the relative size of the trends in $f_r$ with $M_{gr}$
and $M_r$.  In the disk stripping scenario, the change in $f_r$
across the range in $M_{gr}$ should be significantly larger than
the change across the range in $M_r$, as is observed.}
\label{alpha}
\end{figure}

In Figures~\ref{red fraction}a-c, the fraction of galaxies that
belong to the red sequence, $f_r$, is plotted against the r-band
absolute magnitude, $M_r$. In Figures~\ref{red fraction}d-e the
fraction of galaxies with $n_i>2.5$, $f_{n2.5}$, is plotted.
Figures~\ref{red fraction}a,d include all galaxies,~\ref{red
fraction}b(c) includes only galaxies with $g-r<0.7$($g-r>0.7$),
and~\ref{red fraction}e(f) includes galaxies with
$n_i<2.5$($n_i>2.5$). Figure~\ref{red fraction difference}
consists of the same set of plots, only for the difference between
$f_r$ or $f_{n2.5}$ in the groups and in isolation,
$f_{gr}-f_{is}$. In Figure~\ref{break fraction}, similar plots are
made for the fraction of galaxies with large 4000\AA~breaks. The
four lines correspond to isolated galaxies and galaxies in groups
of $log(M_{gr}(M_{\odot}))<~13$, 13 - 14, and 13.5 - 14.5.
Isolated galaxies are defined as galaxies that are either in
singlets or pairs. The $M_r$ bins for Figures 1-4 are chosen such
that the number of galaxies in the middle $M_{gr}$ bin with blue
colors is the same in each bin. This results in uniform errors
across the $M_r$ bins. The plotted errors in Figures~\ref{red
fraction} and~\ref{break fraction} are given by
$\sigma^2=f(1-f)/N$. These errors are propagated to determine the
errors for Figures~\ref{red fraction difference} and~\ref{alpha}.

The observations presented in Figures~\ref{red fraction}
and~\ref{red fraction difference} agree well with the previous
observations discussed in \S~\ref{intro}. Figures~\ref{red
fraction}a,c and~\ref{red fraction difference}a,c show that both
$f_r$ and $f_{n2.5}$ increase with $M_{gr}$. As seen most clearly
in Figure~\ref{red fraction difference}, the excess of galaxies
with $n_i>2.5$ is smaller than the excess of red galaxies.
Figure~\ref{break fraction} demonstrates that the 4000\AA~break
behaves similarly to the $g-r$ color. These relations are the
group-based counterpart of the observations of~\citet{Kauffmann et
al. 03b}, \citet{Kauffmann et al. 04}, \citet{Blanton et al. 05b},
and \citet{Hogg et al. 03} correlating these properties with local
over-density. The large $f_r$ and $f_{n2.5}$ in the groups will
also increase the correlation function due to the higher
clustering of groups compared to isolated galaxies. The increase
in $f_r$ with $M_{gr}$ still appears Figures~\ref{red fraction}b
and~\ref{red fraction difference}b, which include only galaxies
with $n_i<2.5$. However, in Figures~\ref{red fraction}e and
~\ref{red fraction difference}e, which include only blue galaxies,
a difference in $f_{n2.5}$ between groups of different mass is not
seen. From this it can be concluded that the majority of galaxies
responsible for the excess of $n_i>2.5$ galaxies in the groups are
red, but the galaxies responsible for the excess of red galaxies
do not all have $n_i>2.5$. This asymmetry is reflected in the
observation by~\citet{Blanton et al. 05b} that color is a better
predictor of environment that Sersic index.

The analytic model of ram pressure stripping presented in Paper 1
predicts that $f_r$ should increase as $M_{gr}$ increases and as
$M_{sat}$ decreases.  An increase in $f_r$ as $M_{gr}$ increases
is clearly seen in Figures~\ref{red fraction} and~\ref{red
fraction difference}.  In Figure~\ref{red fraction difference}a, a
decrease in $f_{r,gr}-f_{r,i}$ as $L_{sat}$ increases can also be
seen. While the difference between consecutive $M_r$ bins is
small, for the two higher $M_{gr}$ samples, a clear trend is seen
across the range of $M_r$ and the difference in $f_{r,gr}-f_{r,i}$
between the brightest and dimmest bins is several times the
errors. For galaxies with $n_i<2.5$, $f_{r,gr}-f_{r,i}$ increases
with $M_{gr}$ and, for the highest $M_{gr}$ bin, with $M_r$.  The
trend in $f_{r,gr}-f_{r,i}$ with $M_r$ for all galaxies is
shallower in the middle $M_{gr}$ bin and is not observed in
Figure~\ref{red fraction difference}b. Possible reasons for this
are given below and in \S~\ref{d2}.

The observed differences in $f_r$ in the $n_i<2.5$ sub-population
may not reflect true differences.  Both the measurement errors for
$n_i$ and the intrinsic scatter in the two $n_i$ modes are
substantial. Galaxies that scatter out of the $n_i>2.5$
populations into the $n_i<2.5$ population have a higher red
fraction than those scattering in the opposite direction, which
tends to make the measured $f_{r}(n_i<2.5)$, $f_{rm<}$, greater
than the true value, $f_{rt<}$. A consequence of this is that a
change in $f_{n2.5}$ can cause a change in $f_{rm<}$ while
$f_{rt<}$ remains unchanged. In order to examine this effect and
to determine the significance of the observed changes in
$f_{r}(n_i<2.5)$ the following definitions are made: let $n$ be
the percentage of the true $n_i>2.5$ galaxies that scatter into
the observed $n_i<2.5$ population, $t$ be the percentage of the
true $n_i<2.5$ galaxies that scatter in the opposite direction,
and $f_{rt<(>)}$ be the true $f_r$ in the $n_i<(>)2.5$ population.
At fixed $M_r$, $n$ and $t$ should not change with $M_{gr}$.  With
these definitions
\begin{equation}\label{eqn1}
f_{rm<}=\frac{N_{rm}}{N_{m}} =
\frac{f_{rt<}(1-f_{n2.5})(1-t)+f_{rt>}f_{n2.5}n}{(1-f_{n2.5})(1-t)+f_{n2.5}n},
\end{equation}
and for small $\Delta f_{n2.5}$,
\begin{equation}\label{eqn2}
\Delta f_{rm<} = \Delta f_{n2.5}
\frac{(f_{rt>}-f_{rt<})}{(1-f_{n2.5})^2} \frac{n(1-t)}{(1-t+n
f_{n2.5}/(1-f_{n2.5})) ^2}.
\end{equation}
A fortunate property of equation~(\ref{eqn2}) is that for $0<n<1$
and $0<t<1$, $n(1-t)/(1-t+n)^2<0.25$, which is a useful bound when
$f_{n2.5}\sim0.5$. To make a conservative estimate of the size of
$\Delta f_{rm<}$ that can be caused by the increase in $f_{n2.5}$
with $M_{gr}$ in the dimmest $M_{r}$ bin, let $f_{rt>}=1$,
$f_{rt<}=0.4$, $f_{n2.5}=0.45$, $\Delta f_{n2.5} = 0.1$, and
$n(1-t)/(1-t+n)^2=0.25$ . With these values, $\Delta
f_{rm<}\leq0.05$, which is significantly smaller than the observed
change in $f_{r}(n_i<2.5)$ between the low and high $M_{gr}$ bins.
Estimating the size of the induced change in $f_{rm<}$ with $M_r$
is less straight forward because $f_{n2.5}$, $n$, and $t$ all
change. For the highest $M_{gr}$ bin, $f_{n2.5}$ increases from
$\sim0.5$ to $\sim0.8$ across the $M_r$ range. Again assuming that
$f_{rt>}=1$, the induced change in $f_{rm<}$ is
\begin{equation}
\Delta f_{rm<} = \frac{f_{rt<}(1-t_2)+4n_2}{1-t_2+4n_2}
-\frac{f_{rt<}(1-t_1)+n_1}{1-t_1+n_1}.
\end{equation}
In general, scatter between the populations tends to make
$f_{rm<}$ behave similarly to $f_{n2.5}$.  Contrary to this
expectation, while $f_{n2.5}$ increases, $f_{rm<}$ decreases from
$\sim0.6$ to $\sim0.5$. If the scatter between the two populations
decreases as $f_{n2}$ increases, $f_{rm<}$ may decrease. However,
producing the $f_{rm<}$ and $f_{n2.5}$ observed without a decrease
in $f_{rt<}$ requires a somewhat unrealistic decrease in $n$.
Assuming that both $n$ and $t$ are between 0.05 and 0.5, $n$ would
need to decrease by at least 70\% (for example, $n_1=0.5$ and
$n_2=0.15$ or $n_1=0.25$ and $n_2=0.05$ can reproduce the
observations with appropriate values of $t_1$ and $t_2$). It is
therefore probable that both of the observed trends in
$f_{r}(n_i<2.5)$ are real.  The decrease in $f_{n2.5}$ with $M_r$
may contribute to erasing the shallow slope of the correlation of
$f_{r,gr}-f_{r,i}$ with $M_r$ between Figures~\ref{red fraction
difference}a and \ref{red fraction difference}b.

The model's predictions refer specifically to the fraction of the
in-falling blue disk galaxies that eventually join the red
sequence rather than to the difference in $f_r$ between the group
and isolated galaxies. In Figure~\ref{alpha} the difference
between $f_r$ in the groups and in isolation is normalized by the
blue fraction in isolation. This normalized difference is an
estimate of the fraction of the in-falling blue galaxies that join
the red sequence. It is a rough estimate because the properties of
the isolated galaxies may not reflect those of the galaxies that
actually joined the groups. The normalized difference increases
with $M_{gr}$ and for the higher $M_{gr}$ bins decreases with
$L_{sat}$, and the increase with $M_{gr}$ is smaller than that
with $L_{sat}$. According to the model of ram pressure stripping,
the small size of the change in the normalized $f_r$ with $M_r$
results from the comparable sizes of the scatter in $M_{eff}$ and
the range of $M_{sat}$ in the sample. Rather than an indication
that ram pressure stripping is not occurring, the existence of a
small change with $M_r$ is expected if stripping is occurring.

\subsection{Timescale of the SFR Decline}

\begin{figure}[t]
\plotone{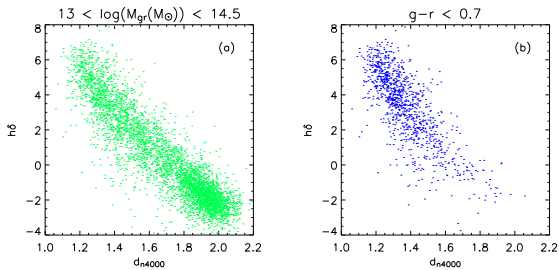} \caption{h$\delta$ vs 4000\AA~break
measure. (a): All galaxies in groups with
$13<log(M_{gr}(M_{\odot}))<14.5$. (b): Galaxies with $g-r<0.7$ for
the same $M_{gr}$. Galaxies that have had their star formation
truncated should reside in the upper-right corner of these plots.
The lack of an excess of galaxies in this region, compared to
isolated galaxies, indicates that color transformation in the
groups is slow.}\label{h delta vs break}
\end{figure}

\begin{figure}[t]
\plotone{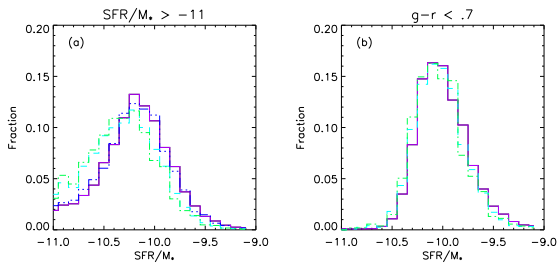} \caption{Normalized $SFR$ distribution in
units of $log(SFR(M_{\odot}/yr)/M_{\ast}(M_{\odot}))$. Lines are
the same as in Figure~\ref{red fraction}. \textit{Left}:
Distribution for all galaxies with $SFR/M_{\ast}>-11$.
\textit{Right}: Distribution for galaxies with $g-r<0.7$. A shift
in the $SFR/M_{\ast}$ distribution for the two high $M_{gr}$ bins
is seen in both plots, particularly in the left panel.  The shift
demonstrates that a population of star forming, blue galaxies with
reduced star formation rates exists in the groups, and is another
indication that SFRs in the groups are declining slowly, as is
expected if stripping of the outer H~\textsc{i} disk is driving
color and SFR evolution in the groups.}\label{SFR hist}
\end{figure}

As discussed in \S~\ref{stripping and sfr}, the timescale over
which a galaxy's SFR should decline after the outer H~\textsc{i}
disk is stripped is of order a Gyr. Evolution of the SFRs on this
timescale is consistent with the galaxies' spectral indices
remaining on the standard track in $H\delta$ versus the
4000\AA~break, as seen in~\citet{Kauffmann et al. 04}.  However,
it is inconsistent with lack of an observable shift in the SFR
distribution for star forming galaxies seen in~\citet{Balogh et
al. 04}.  Both of these tests are repeated here to determine
whether color evolution in these groups is fast or slow.

Figure~\ref{h delta vs break} shows $H\delta$ versus the
4000\AA~break for $13>log(M_{gr}(M_{\odot}))<14.5$ for all
galaxies and for galaxies with $g-r<0.7$. Compared to Figure 8
in~\citet{Kauffmann et al. 04}, there is no excess of galaxies
residing above the standard track, where a galaxy that has had its
star formation abruptly truncated should lie.   In Figure~\ref{SFR
hist} distributions of $SFR/M_{\ast}$ are plotted. Figure~\ref{SFR
hist}a shows the distributions for star-forming galaxies,
$log(SFR/M_{\ast})>-11$, and Figure~\ref{SFR hist}b shows the
distributions for galaxies with $g-r<0.7$. For the medium and
large $M_{gr}$ bins, a clear shift in the distribution of
$SFR/M_{\ast}$ is seen for the star-forming galaxies. A KS test
gives a less than 0.01\% probability that the $SFR/M_{\ast}$
distributions in the groups and in isolation are drawn from the
same sample. Though less visually apparent, a slight shift also
exists for the blue galaxies. The average $SFR/M_{\ast}$ is
slightly lower in the groups than in isolation, and a KS test
gives a 0.5\% probability that they are drawn from the same
distribution.  These observations are consistent with a decline in
the star formation rate on timescales of order a Gyr.

In Figure~\ref{SFR hist}, the $SFR/M_{\ast}$ values
from~\citet{Brinchmann et al} are used rather than the $H\alpha$
equivalent widths, $W_0(H\alpha)$, which are used in~\citet{Balogh
et al. 04}. The $SFR/M_{\ast}$ are likely a better measure of the
relative star formation rate in part because they measure
$SFR/M_{\ast}$ rather than $SFR/L$.  The $SFR$ distributions
presented here also differ from the~\citet{Balogh et al. 04}
distributions in that group membership is used for all
comparisons. When the $W_0(H\alpha)$ distributions for all
isolated and group galaxies with $W_0(H\alpha)>4$\AA~are compared,
$\langle W_0(H\alpha)\rangle$ is lower for the group galaxies and
a KS test rules out the possibility that the two populations are
drawn from the same distribution. This results disappears when
only galaxies with $W_0(H\alpha)>10$\AA~are considered.  However,
if the galaxies' $W_0(H\alpha)$ are divided by their
$M_{\ast}/L_z$ ratios, as measured by~\cite{Kauffmann et al. 03a},
then the difference in the $W_0(H\alpha)$ distributions remain
regardless of the cut used. In both instances a slight difference
is also seen in the blue population, and a KS test gives a few
percent chance that the blue group and isolated galaxies are drawn
from the same population.

The size of any shift in the distribution of $SFR/M_{\ast}$ in the
blue galaxy population is likely to be small.  As can be seen in
Figure~\ref{alpha}, approximately half of the in-falling blue
galaxies are unaffected by reddening.  The $SFR/M_{\ast}$
distribution for blue galaxies is therefore a superposition of the
unaffected blue galaxy population and a transiting population of
blue galaxies with low SFRs that have not yet joined the red
sequence. In addition, the difference in $SFR/M_{\ast}$ between an
average in-falling blue galaxy, $\log(SFR/M_{\ast})\approx-10$,
and the $SFR/M_{\ast}$ at which galaxies appear to leave the blue
population, $\log(SFR/M_{\ast})\approx-10.4$, is small.
Furthermore, the use of color to estimate the effect of the finite
fiber size on $SFR/M_{\ast}$ biases against seeing a shift in
$SFR/M_{\ast}$ in the blue galaxy population. The change in the
average $\log(SFR/M_{\ast})$ in the middle(high) $M_{gr}$ bin is
0.04(0.06).  If the decay in the SFR is exponential, then placing
20\%(30\%) of the blue galaxies halfway along a journey from their
incoming $SFR/M_{\ast}$ to joining the red sequence will result in
this shift.  Therefore, a significantly larger shift in the
$SFR/M_{\ast}$ than that observed for the blue group galaxies
isn't expected.

The location of the galaxies in this sample in the H$\delta$
versus 4000\AA~break plane and the relative distributions of
$SFR/M_{\ast}$ for star-forming and blue galaxies in the groups
versus in isolation both indicate that galaxies that transition
between the blue and red populations experience a decline in their
SFRs on long timescales. This is consistent with color evolution
that is driven by the ram pressure stripping of gas from the outer
H~\textsc{i} disks of blue disk galaxies.

\subsection{The Larger Picture}\label{d2}
The differences observed in color and structure between the group
and isolated galaxy populations can be summarized as follows.  For
both galaxies with $n_i>2.5$ and $n_i<2.5$, the red fraction is
larger in the groups than in isolation. This can be seen in
Figures~\ref{red fraction}b,c and~\ref{red fraction
difference}b,c. In addition, color selected galaxy populations
have identical $n_i$ distributions both inside and outside groups
(Fig.~\ref{red fraction}e,f \&~\ref{red fraction difference}e,f;
\citep{Quintero}). In this section, the role that ram pressure
stripping may play in this larger picture is discussed.

In isolation, $f_{n2.5}$ for red galaxies and $f_r$ for galaxies
with $n_i<2.5$ are both large. Therefore, for the color selected
$n_i$ distributions in the groups to match those in isolation, the
normalized difference in the red fraction between group and
isolated galaxies, that is $(f_{r,gr}-f_{r,i})/(1-f_{r,i})$, must
be larger for galaxies with $n_i>2.5$ than for galaxies with
$n_i<2.5$. In addition, though the $n_i$ distributions within
color selected subsamples are identical in the group and isolated
galaxy populations, the larger red fraction in the groups is
accompanied by a larger fraction of galaxies with $n_i>2.5$. Ram
pressure stripping can create an excess of red galaxies in groups
in both the $n_i>2.5$ and $n_i<2.5$ populations.  In particular,
because ram pressure stripping does not affect galaxies'
structures, it can transform a blue disk galaxy into a red disk.
Furthermore, a greater fraction of the in-falling spherical
galaxies that have gas will be stripped than of the in-falling
disk galaxies. However, ram pressure stripping cannot account for
the structural differences that are observed.

It is worth pointing out that, while in the most massive groups in
this sample almost no mergers should occur, the merger rates  in
the low and medium-mass groups may be quite high~\citep{Mamon}. In
addition, N-body simulations show that a greater fraction of the
halos that reside in clusters today underwent major mergers in
their pasts~\citep{GKK}.  Therefore, the high-mass group galaxy
population may have experienced an enhanced merger rate in the
past while the low- and middle-mass group galaxy populations may
be experiencing enhanced merger rates currently.  In this light,
Figure~\ref{red fraction difference}b becomes interesting. The
slope of the correlation between $f_{r,gr}-f_{r,i}$ and $M_r$ is
shallower in the middle $M_{gr}$ bin than in the high $M_{gr}$
bin. Mergers may tend to wash out the dependence of $f_r$ on
$M_{sat}$ in the middle-mass groups while the dependence is
preserved in the high-mass groups. A larger merger remnant
fraction in the groups may, of course, also account for some of
the structural differences observed in the group galaxy
populations.

\section{CONCLUSIONS}\label{conclusion}
Observations of the red fraction, $f_r$, the fraction of galaxies
with a large 4000\AA~break, $f_{4000}$, and the fraction of
galaxies with $n_i>2.5$, $f_{n2.5}$, in the SDSS groups are
presented in Figures~\ref{red fraction} and~\ref{red fraction
difference}. These are the group-based analog of previous
observations and are consistent with the correlations between
galaxy properties and local environment in~\citet{Kauffmann et al.
03b, Kauffmann et al. 04}, \citet{Hogg et al. 03}, and
\citet{Blanton et al. 05b}. In addition, the observed excess of
red galaxies in the groups regardless of $n_i$ makes color a
better predictor of environment, as observed in~\citet{Blanton et
al. 05b}.

In \S~\ref{Model} several predictions are made for the group
sample using an analytical model of ram pressure stripping
presented in~\citet{P1}. In general, the model predicts that the
red fraction should increase as either the group mass, $M_{gr}$,
increases or the satellite mass, $M_{sat}$, decreases.  For the
SDSS group sample in particular, it predicts that mild stripping
should occur in the low-mass groups while moderate to severe
stripping should occur in the middle- and high-mass groups.
Therefore, differences in $f_r$ with both $M_{gr}$ and $M_r$ are
expected if ram pressure stripping is driving color change. The
expected scatter in an effective mass for stripping at fixed
$M_{sat}$ is smaller than the range of $M_{gr}$ present in the
sample, but comparable to the range of $M_{sat}$.  Therefore, in
the stripping scenario, observed changes in $f_r$ with $M_r$
should be small while the change with $M_{gr}$ should be larger.
In addition, if stripping of the outer H~\textsc{i} disk, along
with the hot galactic halo, is causing the decline in SFRs in the
groups, then the timescale of the decline should be of order a
Gyr.

The predicted behaviors are observed in the group sample.  As seen
in Figures~\ref{red fraction}ab,~\ref{red fraction difference}ab,
and~\ref{alpha}, the red fraction increases with both $M_{gr}$
and, for the higher mass groups, with $M_{r}$ for both all
galaxies and galaxies with $n_i<2.5$. This is despite a decrease
in $f_{n2.5}$ with $M_r$. In the lowest mass groups, where only
mild disk stripping is expected, the increase in $f_r$ is slight
and may be due to galactic halo stripping alone. In
Figure~\ref{alpha}, an estimate of the fraction of the in-falling
blue galaxies that join in the red sequence after entering a group
is presented.  This estimate is the best test of the model's
predictions for the sizes of the change in $f_r$.  As predicted,
the change in $f_r$ across the $M_r$ range is quite small while
the change across the $M_{gr}$ range is significantly larger. As
shown in Figures~\ref{h delta vs break} and~\ref{SFR hist}, the
positions of the group members in the H$\delta$ versus
4000\AA~break plane and the distributions of $SFR/M_{\ast}$ for
the star forming and blue galaxies both demonstrate that SFRs in
the groups are declining on timescales greater than a Gyr. These
observations all indicate that the observed color differences in
the groups are the result of the ram pressure stripping of the
outer H~\textsc{i} disk followed by a gradual decline in the
galaxies' SFRs.

One alternate scenario for color evolution in groups postulates
that the two dominate processes in groups are galactic halo
stripping and mergers. The observed trends in Figures~\ref{red
fraction} and~\ref{red fraction difference} do combine color
change for galaxies at all values of $n_i$ and structural
differences which are accompanied by color differences. If a
galaxy's structure is altered, for instance by a major merger,
while it orbits in a group, then any remaining gas should be
quickly stripped and the SFR should decline.  In addition, if star
formation in disk galaxies is fed only by the cooling of the
galactic halo directly onto the star-forming disk, then the
stripping of this halo will result in red galaxies at all $M_r$
and $M_{gr}$. Similar uniform reddening should be seen for
spherical galaxies.

The observed trends in $f_r$ in Figures~\ref{red fraction}ab,
\ref{red fraction difference}ab, and \ref{alpha} are an excellent
match to predictions based on the ram pressure stripping model,
which is in itself a strong indication that disk stripping plays a
role in galaxy evolution in groups.  However, the observation that
the red fraction increases with group mass is particularly strong
and can be used to differentiate between the scenario presented in
the previous paragraph and the disk stripping scenario. In
\S~\ref{Model}, the model presented in Paper 1 is combined with an
estimate of the maximum gas density that can be sustained in these
galaxies in order to place an upper limit on the fraction of the
hot halo gas that the galaxies can retain.  All galaxies in both
the middle- and high-mass groups should be stripped of the
majority of their galactic halo gas.  Considering the small mass
of the remaining gas and the absence of any new in-falling gas,
the remaining galactic halo cannot continue to feed star-formation
in any of these galaxies. This would lead to the expectation that
$f_r$ should not depend on either $M_{gr}$ or $M_{r}$ in these
groups.  This loose theoretical expectation is affirmed by
simulations. In N-body based semi-analytic models in which star
formation is shut off when satellite galaxies enter a group, the
red fraction in groups is independent of
$M_{gr}$~\citep{Weinmann06}. However, in Figure~\ref{red
fraction}b a substantial difference in $f_r$ is seen for groups of
different $M_{gr}$. This difference, which is unexpected and must
be rationalized in the strangulation scenario, is easily
understood in the disk stripping scenario.

The observations presented in this paper favor ram pressure
stripping of the outer H~\textsc{i} disk as the dominant driver of
color change for disk galaxies in groups.  Despite this, other
processes must occur at some level. Galaxies in groups are
expected to undergo mergers and some merging galaxies will have
retained their H~\textsc{i} disks.  In the dense inner regions of
clusters stripping of gas from within the stellar disk should
occur on occasion.  Finally, even galaxies that retain their disks
are stripped of their galactic halo and will eventually experience
a decline in the SFRs. However, for the low redshift groups
studied here, ram pressure stripping appears to be driving the
relationships between SFR, color, $M_r$, and environment.

Sorting out which processes are dominant for different $M_{sat}$
and at different redshifts will require combining further modeling
and observations. To determine the importance of ram pressure
stripping it is most important to understand how galaxies fuel
their star formation.  Understanding trends like those shown in
Figures~\ref{red fraction} and~\ref{red fraction difference} will
also require understanding the processes that can alter structure.

\section{ACKNOWLEDGEMENTS}
This project was advised by D. N. Spergel and funded by NASA Grant
Award \#NNG04GK55G.  I'd like to thank A. Berlind, M. Blanton, and
D. Hogg for the use of the SDSS group catalog and for useful
discussions.

Funding for the Sloan Digital Sky Survey (SDSS) has been provided
by the Alfred P. Sloan Foundation, the Participating Institutions,
the National Aeronautics and Space Administration, the National
Science Foundation, the U.S. Department of Energy, the Japanese
Monbukagakusho, and the Max Planck Society. The SDSS Web site is
http://www.sdss.org/.

The SDSS is managed by the Astrophysical Research Consortium (ARC)
for the Participating Institutions. The Participating Institutions
are The University of Chicago, Fermilab, the Institute for
Advanced Study, the Japan Participation Group, The Johns Hopkins
University, the Korean Scientist Group, Los Alamos National
Laboratory, the Max-Planck-Institute for Astronomy (MPIA), the
Max-Planck-Institute for Astrophysics (MPA), New Mexico State
University, University of Pittsburgh, University of Portsmouth,
Princeton University, the United States Naval Observatory, and the
University of Washington.

\end{document}